\documentclass{aastex}

\tighten

\def\lsim{\lower.5ex\hbox{$\; \buildrel < \over \sim \;$}}
\def\gsim{\lower.5ex\hbox{$\; \buildrel > \over \sim \;$}}
\def\lax {\ifmmode{_<\atop^{\sim}}\else{${_<\atop^{\sim}}$}\fi}
\def\gax {\ifmmode{_>\atop^{\sim}}\else{${_>\atop^{\sim}}$}\fi}
\def\etal{{\it et al.\/} }
\def\gtorder{\mathrel{\raise.3ex\hbox{$>$}\mkern-14mu
\lower0.6ex\hbox{$\sim$}}}
\def\ltorder{\mathrel{\raise.3ex\hbox{$<$}\mkern-14mu
\lower0.6ex\hbox{$\sim$}}}

\def\pmb#1{\setbox0=\hbox{#1}%
\kern-0.015em\copy0\kern-\wd0
\kern0.03em\copy0\kern-\wd0
\kern-0.015em\raise0.0433em\box0 }

\begin{document}

\title{Normal Branch Quasi-Periodic Oscillations in Sco X-1: Viscous
Oscillations of a Spherical Shell Near the Neutron Star}

\author{L. G. Titarchuk}
\affil{George Mason University/Center for Earth Observing and Space
Research, Fairfax VA 22030}
\affil{US Naval Research Laboratory, Washington, DC 20345-5352}
\affil{Laboratory for High Energy Astrophysics, Goddard Space Flight
Center, Greenbelt, MD 20771}
\email{lev@xip.nrl.navy.mil}

\author{C. F. Bradshaw \& B. J. Geldzahler}
\affil{School of Computational Sciences, George Mason University, Fairfax
VA} \email{cbradsha@qwestinternet.net \& bgeldzahler@hotmail.com
<mailto:bgeldzahler@hotmail.com>
<<mailto:bgeldzahler@hotmail.com>>}

\author{E. B. Fomalont}
\affil{National Radio Astronomy Observatories, Charlottesville VA }
\email{efomalon@nrao.edu}
\vskip 0.5 truecm


\begin{abstract}

We present a comprehensive classification of all
observed QPOs within the framework of the transition layer model using a
large set of Rossi X-ray Timing Explorer (RXTE) data for Sco X-1.
The model assumes an optically thin material along
the observer's line of sight in the horizontal branch and an increasingly
optically thick material while in the other two branches that is
consistent with X-ray and radio observations and the disk transition layer
model of QPOs. We identify the $\sim 6$ Hz frequencies in the normal
branch as acoustic oscillations of a spherical shell around the neutron
star (NS) that is formed after radiation pressure near the Eddington
accretion rate destroys the disk. The size of the shell is on the order of
one NS radii from the NS. 
We also estimate the upper limit of Sco X-1's magnetic field
to be $0.7 \times 10^6$G at about one NS radii above the NS surface while
in the horizontal X-ray branch.

\end{abstract}


\section{Introduction}
%
%

Sco X-1 is the brightest persistent non-solar X-ray
source in the sky (Giaconni \etal 1962). Sco X-1's distance is $2.8 \pm
0.3$ parsec (Bradshaw \etal 1997) and its X-ray luminosity is close to the
theoretical Eddington limit for a 1.4 $M_{\odot}$ neutron star
($\sim 2 \times 10^{38} \: erg \: s^{-1}$). Sco X-1 belongs to a class of
'Z'-type low mass X-ray binaries (LMXBs) that show a Z-shape in their
X-ray color diagram (CD) and their hardness-intensity diagram (HID)
(Hasinger \& van der Klis 1989). The upper branch of this diagram is
called the horizontal branch (HB), the lower branch is called the
flaring branch (FB), and the normal branch (NB) connects the other two
to form the Z. All of the Z-type sources have a similar-shaped FB and
NB, but not HB. The Cygnus-like sources (Cyg X-2, GX5-1 and GX 340+0)
exhibit a ``horizontal'' HB, whereas the Sco-like sources
(Sco X-1, GX 17+2 and GX 349+2) exhibit a
``vertical'' HB (Kuulkers \etal 1997). The differences in the two Z
source groups have been suggested to result from different inclination
angles to the line-of-sight (Hasinger \& van der Klis 1989; Hasinger \etal
1989; Hasinger \etal 1990a; Kuulkers \etal 1994; Kuulkers 1995) or
differing NS magnetic field strength (Psaltis \etal 1995).

Most Z-sources also exhibit X-ray quasi-periodic oscillations (QPOs) from
a few hertz to kilohertz which vary in intensity and frequency as a
function of position in the X-ray CD. The X-ray NB is characterized by
the presense of 1\%-5\% amplitude, 6-20 Hz normal/flaring-branch
quasi-periodic oscillations (N/FBOs). There is an almost linear
correlation between Sco X-1's kHz and 6 Hz QPOs.  The standard interpretation (see van
der Klis 1995 and references therein and Hasinger \etal 1990b) of the 6 Hz
QPO is that, as the mass accretion rate increases to near Eddington, part of
the accretion occurs in an approximately spherically symmetric radial inflow with a
radius of a $\sim$100 km. 

The standard interpretation was proposed before the discovery of kHz QPOs
(Strohmayer \etal 1996; van der Klis \etal 1996; Zhang \etal 1996)
and could not discuss any qualitative and quantitative dependence between
these two QPO features. A new explanation is needed because it was
never clear how frequencies produced by structures/phenomena
with sizes of $\sim 100$ km could be strongly correlated with kHz QPO
frequencies, presumably occuring in regions with sizes of an order of
magnitude less ($\sim 15$ km).

Titarchuk, Lapidus \& Muslimov (1998), hereafter TLM98,
considered the possibility of the dynamical adjustment of a Keplerian disk
to the innermost sub-Keplerian boundary conditions (e.g., at the surface
of a NS) to explain most of Sco X-1's observed QPOs [transition layer
model (TLM)]. In TLM98 the authors conclude that a transition
sub-Keplerian layer between the NS surface and the last Keplerian orbit forms as a
result of this adjustment. We have extended the TLM98 analysis to include the $\sim
6$ Hz low frequency oscillations observed in the NB.

In \S 2, we briefly describe the new RXTE and radio observations of Sco
X-1. After a summary of the variety of QPO phenomenon in Z-sources, we
confirm the various QPO relationships of TLM and the QPO classification
of Titarchuk, Osherovich \& Kuznetsov (1999) hereafter TOK99, 
and of  Wu (2001) to include
6 Hz NBOs in \S 3. In \S 4, armed with the RXTE observation and the
orientation to the line-of-sight of Sco X-1, we suggest that the strong
6 Hz N/FBOs are associated with viscous oscillations in an accreting
spherical halo surrounding the NS. In \S 5, we calculate the characteristic velocities of
the transition layer from observed values of the viscous, and kHz
QPO frequencies and finally, we estimate Sco X-1's magnetic
field based on observational data. Our summary and conclusions
follow in \S 6.

\vskip 0.25 truecm
\centerline{\bf 2. Observations}

We observed Sco X-1 as part of a multi-frequency X-ray and radio
(Fomalont et al 2001a \& b) campaign using the RXTE proportional counting
array (PCA) during August 3 and 22, 1997, February 27 and 28, 1998, and June
10-13, 1999. Data were simultaneously collected in the 2-60 keV energy
band at both 16s and 250$\mu$s resolutions. The 250$\mu$s data produced power density
spectra for determining QPOs. The power density spectra were fitted with
either a constant plus a Lorentizan or a power law plus a Lorentizan
to determine QPO frequencies. Details of these observations will be
provided in Bradshaw \etal (2002).

\vskip 0.25 truecm
\centerline{\bf 3. Classification of Sco X-1's QPOs, Including 6 Hz NBOs}

Understanding QPO relationships allows us to develop a better
understanding of the accretion processes and physical parameters in bright
LMXBs. TOK99 suggested a ``QPO classification'' for Sco X-1's observed
frequencies that includes:
(1) the Keplerian frequency of matter above a transition layer, $\nu_{\rm K}$,
(2) a hybrid frequency $\nu_h$, greater in frequency than $\nu_{\rm K}$,
(3) and a horizontal branch oscillation (HBO), $\nu_L$,
both of them are formed due to the Coriolis force in the rotational frame
of reference, (4) the second harmonic, 2$\nu_L$ of the HBO, discovered by van der
Klis et al. (1996), (5) a viscous frequency, $\nu_V$, related to accreting matter's
radial drift timescale, and (6) a break frequency $\nu_b$, related to the radial diffusion
timescale in the transition layer.

\begin{figure}
\epsscale{0.75}
\plotone{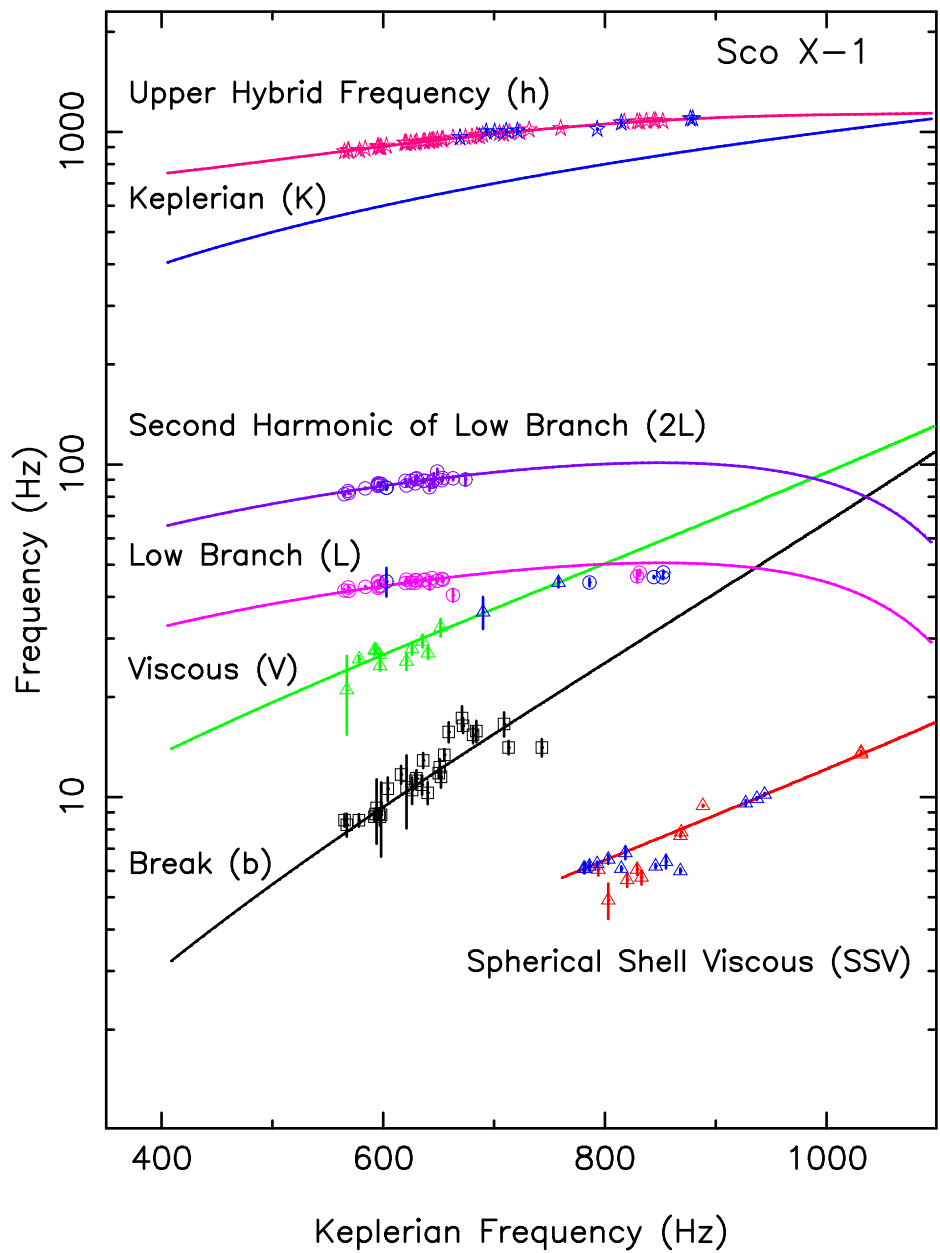}
\caption{Classification of QPO branches 
(hybrid $\nu_h$, Keplerian  $\nu_{\rm K}$, low branch $\nu_L$ and
$2\nu_L$, viscous $\nu_V$, break $\nu_b$ and spherical shell viscous
$\nu_{ssv}$) in the Z source Sco X-1. This
classification is an extend version of the
QPO classification for Sco X-1 in TOK99. 
The plot includes QPOs from our 1997, 1998, and 1999 RXTE
observations (blue star, circle, and triangle symbols) and 6 Hz
observations that represent viscous frequencies under spherical
accretion.}
\label{fig1}
\end{figure}

Correlations between kHz frequencies and 6 Hz (NBO) frequencies were
presented in Figure 4 and Table 1 of van der Klis \etal (1996, hereafter
VDK96) and correlations between kHz frequencies and HBO frequencies were
presented in van der Klis \etal (1997, hereafter VDK97).
The VDK96,97 data were obtained by the RXTE during observations
of Sco X-1 in February 1996. TOK99 compared those observations with a
theoretical dependence derived using the TLM, which is
included in Figure \ref{fig1}. VDK96,97 and our 1997, 1998, and 1999
observations found upper kHz frequencies for almost all
observations of Sco X-1. However, low kHz (Keplerian) frequencies were
only detected in about half of the observations, requiring us to
convert upper kHz frequencies to Keplerian frequencies using the upper
hybrid relation defined in Osherovich \& Titarchuk (1999) to
compare Keplerian and 6Hz frequencies.
The frequencies of the observed QPOs for Sco X-1 are plotted in Figure
\ref{fig1} as a function of the Keplerian frequency. {\it In addition to
the classification plot presented in TOK99 (Figure 2), we include $\sim 6$
Hz frequencies, representing spherical viscous oscillations, and newly
observed QPO points, marked by blue stars, circles and triangles}.
The 6 Hz QPOs have frequencies satisfying the theoretically derived
dependence of the {\it viscous} frequencies $\nu_V$ on $\nu_{\rm K}$. The theoretical curve
for the $\sim 6$ Hz frequencies is identical to that for the viscous
frequencies but with a normalization coefficient $C_N=1.25$ (see TOK99,
Eq. 6).
{\it We find that the $\sim 6$ Hz frequencies are related to the viscous frequency branch but the
normalization of the theoretical dependence of viscous frequencies on the 
Keplerian frequencies drops almost an order of magnitude (factor of 7.8 for
Sco X-1) when the accretion rate (or luminosity) approaches Eddington}.

\vskip 0.25 truecm
\centerline{\bf 4. 6 Hz Viscous Oscillations of a Spherical Halo}

We suggest that the viscous frequency change by a factor of about an order
of magnitude is caused by a radical change in the cross-section of
the accretion flow. In Titarchuk and Osherovich (1999), hereafter TO99,
the authors argue that $\nu_V\propto v_r/L$ where $v_r$ is a radial component
of the accretion flow velocity and $L$ is a size of the transition layer,
(i.e the distance between the transition layer outer boundary, $R_{out}$
and the NS surface, $R_0$). In the TLM,
$R_{out}$ is an adjustment radius where the Keplerian disk adjusts to the
sub-Keplerian transition layer (TL) and the QPO frequency $\nu_{\rm K}$ is a Keplerian
frequency at $R_{out}$. Thus, a small change of $\nu_{\rm K}$ should lead to a
small change of the transition layer size and, hence, the viscous
frequency. Under these assumptions, the only reason for a break in the viscous frequency is
a break in the radial velocity $v_r$.
Assuming continuity of the accretion flow, we conclude that,
at the stage corresponding to $\nu_{\rm K}$ near 800 Hz (i.e, $R_{out}\sim
2R_0$), the disk-like transition layer no longer exists. The accretion
geometry becomes quasi-spherical down to the NS surface. {\it Hence, while
the mass accretion rate changes relatively little, the geometry
of accretion changes drastically and the cross-section of the flow increases by almost 
an order of magnitude (from $4\pi RH$, $H \ll R$, to $4\pi R^2$)}.

\begin{figure}
\epsscale{0.75}
\plotone{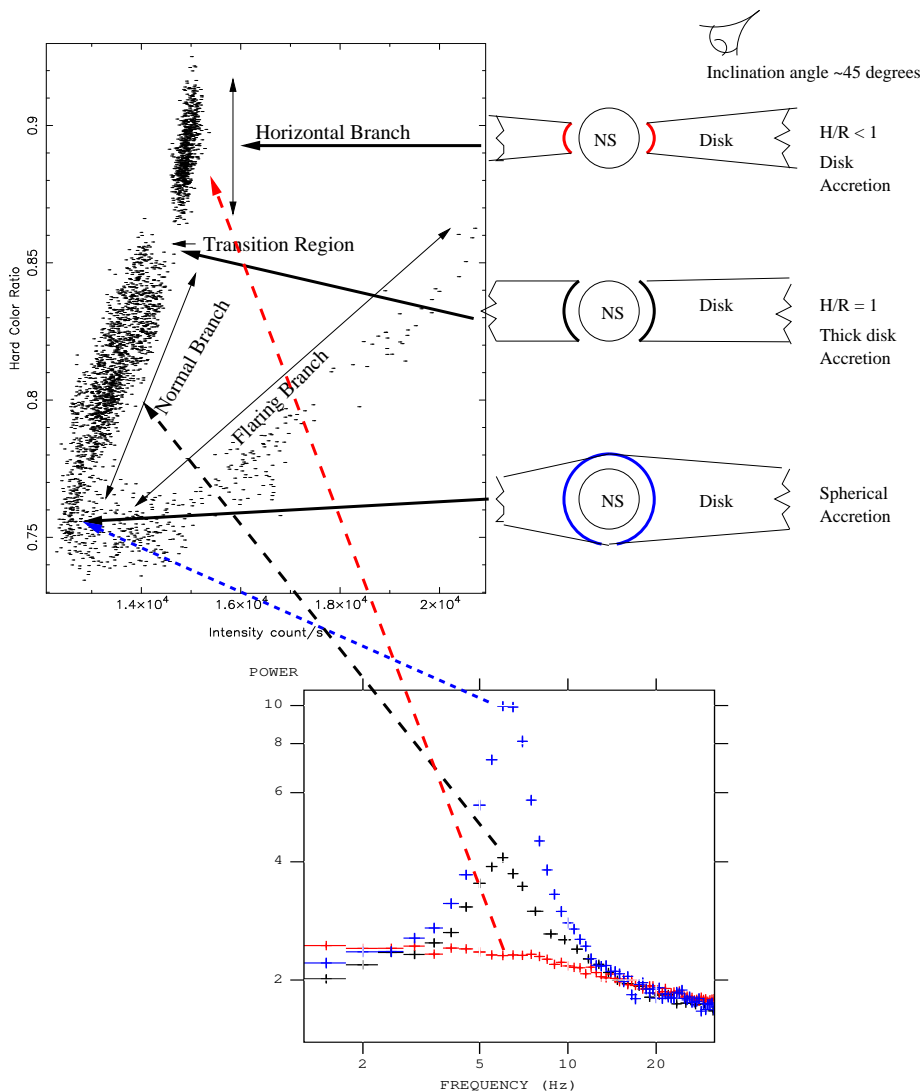}
\caption{The RXTE hardness-intensity diagram of Sco X-1 with an overlay
illustrating hardness ratios producing QPOs.
QPOs near 6 Hz are produced along the normal branch, becoming more
coherent as the inferred accretion rate increases.
The horizontal branch does not have a coherent QPO near 6 Hz, but does
have a break frequency.
The three figures to the right represents the accretion changing from a
disk to spherical accretion.
The intensity reflects background corrected counts from PCU 0 only. The
RMS errors are $\sim$ 3.5\%. The hard color ratio is (9.1-18.2 keV) /
(6.9-9.1 keV). The data were observed by the RXTE during June 10-13, 1999,
ObsId P40706.}
\label{fig2}
\end{figure}

This geometrical change leads to a frequency mode change and approximately
an order of magnitude decrease of the viscous frequency.
The high accretion rate puffs up the transition layer and matter accretes
onto the NS almost spherically, forming a transition shell
instead of a transition layer. This shell produces a frequency [see
Titarchuk, Bradshaw \& Wood (2001), hereafter TBW01 and compare with
Hasinger (1987)]
\begin{equation}
\nu_{ssv} =\frac{fv_s }{L} \: {\rm Hz},
\end{equation}
where $v_s$ is the sonic velocity and $f$ is 0.5 and $\sim 1/2\pi$
for the stiff and free  boundary conditions in TL respectively.
Figure \ref{fig2} provides a schematic of the evolution to spherical
accretion. This conclusion, based on the radical viscous (NBO) frequency change in
the TLM, is similar to that suggested by Hasinger
(1990). However, our model
interprets the $\sim 6$ Hz QPOs as acoustic oscillations of a spherical shell
surface located within 20 km of the NS surface instead of the radial inflow within a
radius of $\sim$100 km, presented in the standard model.

Observational support for a switch between the disk scale height-neutron
star radius ratio from $H/R \ll 1$ to $H/R \sim 1$ in Sco X-1 may be
provided by its vertical HB and by Sco X-1's $\sim 45$ degree look angle (Fomalont \etal
2001a).
The intensity of a photon source embedded in a medium of optical depth
$\tau$ and with the scattering probability $\lambda$ is given by 
$S_{obs} = S_{0} \exp({- k\tau})$ (e.g. Sobolev 1975),
where $k\tau= [3(1-\lambda)]^{1/2} \tau_T$ is an effective optical depth
taking into account the electron (Thomson) scattering of X-ray photons and
extinction (e.g., due to free-free and bound-free absorption).
Assuming a flux source function, $S_{obs}$, given by this equation
 a balance is achievable between the variable intrinsic flux
($S_{0}$) and the variable optical depth that results in an approximately 
constant observed intensity ($S_{obs}$) in HBO.

The HB is presumed to have the
least accretion rate and above the transition region noted in Figure
\ref{fig2} (disk accretion), smaller amounts of accreting material obscure the
observer's line-of-sight to the NS.  As the accretion rate increases sufficiently to
produce a disk scale height to NS radius $H/R \sim 1$, more matter obscures the observer's
line-of-sight to the NS and Sco X-1 enters the NB.
Further increases in accretion rate linearly increase the source
intensity, but the observed optical depth increases too, from both a
relative density increase and an increase in the path length through the obscuring
material. This results in the decrease in observed intensity (because of
the exponential dependence of $S_{obs}$ on $\tau$) and a
softening of the X-ray flux, as observed in the normal branch of Sco X-1's
HID and CD. The Sco-like source, GX 17+2, has a HID and transition
region gap (Wijnands \etal 1997, Figure 1b) almost identical to Sco
X-1. GX 17+2's horizontal branch HID shows a slight slope toward lower
intensity as the hard-color ratio increases. In the context of our model,
GX 17+2 has a line-of-sight inclination angle that is greater, by a small
amount, than the inclination angle of Sco X-1.

For a Cygnus-like Z source, the look angle is presumably
greater than Sco X-1's and thus the line-of-sight is nearly tangent to the
disk surface. The X-ray emission region is effectively semi-infinite in this case, i.e.
the observed radiation flux is more independent of the emission region's
optical depth and $S_{obs}$ is linearly proportional to the mass accretion rate.
As the accretion rate decreases past a point where the disk scale
height is $\ltorder 1$, $S_{obs}$ decreases linearly with accretion rate, resulting in a
HB with hardness increasing slowly and intensity decreasing rapidly as seen
in Cyg X-2's HID and CD. 

\vskip 0.25 truecm
\centerline{\bf 5. The Transition Layer B-field}
\vskip 0.25 truecm
\centerline{\bf 5.1 Viscous Velocity}

Solutions presented in TO99 showed that, 
we can interpret $\nu_V$ as a frequency of viscous (presumably magnetoacoustic)
oscillations in the transition region.
The fundamental frequency of acoustic oscillations in a bounded
medium with the free boundary conditions is 
 $\nu_V\approx v_{aco}/2\pi L$ (TBW01, see also Eq. 1).
For a given Keplerian frequency 
$\nu_{\rm K}$ the transition layer size $L =R_{out}-R_0$,
where $R_{out}=[GM/(2\pi\nu_{\rm K})^2]^{1/3}$ and $R_0=3R_{\rm S}=6GM/c^2$
and $G$ is the gravitational constant.
Thus, $v_{aco}\approx 6L\nu_{V}$
and for  given $\nu_V=32.4$ Hz, $\nu_{\rm K}=652$ Hz (TOK99) and
$M=1.4M_{\odot}$, $L=9.8\times10^5$ cm, $v_r\approx
v_{aco}=1.9\times10^{8} \: cm \: s^{-1}$.
This radial velocity, $v_r$, estimate translates to
a plasma temperature of the transition layer of about 19 keV, provided
that the viscous oscillation is determined only by thermal
motion and can be verified by fitting the Sco X-1 photon
spectra. However, our photon spectra fitting resulted in a transition
layer temperature of 2.8 keV, suggesting that the oscillation is not determined
only by thermal motion.

\vskip 0.25 truecm
\centerline{\bf 5.2 B-field estimate}

We fitted HB and NB spectra between 8.0-25 keV with
the Comptonization model of Sunyaev \& Titarchuk (1980) (compST). The
fitted data yielded a plasma temperature and optical depth of the transition 
layer of $kT=2.8$ keV and $\tau=10.5$,
respectively. From this temperature, we derive a ``Comptonization
velocity'' $v_{\rm comp}\approx(kT/m_p)^{1/2}$ of $5.2 \times 10^{7}\:cm \: s^{-1}$.
This velocity is a factor of 4 smaller than $v_r$.
In order to find the upper limit of the magnetic field of strength $B$ at
the outer radius of the transition layer, we assume that the viscous
frequency is determined by compresssional Alfv\'en waves only (see, Lang 1999
for a definition of the Alv\'en wave). TBW01 found that for Alfv\'en waves
\begin{equation}
\nu_V\approx 2v_{\rm alf}/R_{out}
\label{eq:5}
\end{equation}
where $v_{\rm alf}=B/(4\pi \rho)^{1/2}$ is the Alfv\'en velocity for
magnetic field strength $B$ at $R_{out}$ and the matter density $\rho$.
We express $\rho=\tau m_p/(\sigma_T L)$.
Thus we can estimate $B$ as
\begin{equation}
B\leq(4\pi\rho)^{1/2}\nu_{\rm V}R_{out}/2\approx(4\pi \tau
m_p/\sigma_T/L)^{1/2}\nu_VR_{out}/2 .
\end{equation}
In \S 5.1, we calculated $L=9.8\times 10^5$ cm using $R_{out}=2.23\times
10^6$ cm as being related to $\nu_{\rm K}=651.7 Hz$. By using $\tau=10.5$
and $\nu_V=32.4$, {\it we estimate the upper limit of the magnetic field
strength to be $B=0.7\times10^6$G at 22 km radius, e.g., the outer edge of
the disk transition layer of Sco X-1 in the HB}. This distance is about
one NS radii from the surface.
It is worth noting that deriving the B-field near the
NS by this approach is based on our observation-based knowledge of the
lower kHz QPO frequency $\nu_{\rm K}$, viscous frequency $\nu_V$, and
best-fit Comptonization optical depth $\tau$ and temperature $kT$.
If a velocity $v_r$ inferred from $\nu_V$ and $\nu_{\rm K}$ is much
greater than $v_{\rm comp}$, we estimate the B-field using $\nu_V$ as a
frequency of the Alv\'en wave.
Thus, this approach is general and applies to any NS system for which the
above information is available.

\vskip 0.25 truecm
\centerline{\bf 6. Summary and Conclusions}
A simple model of Sco X-1 that assumes an optically thin material along
the observer's line of sight in the horizontal branch and an increasingly
optically thick material while in the other two branches is consistent
with X-ray and radio observations and the disk transition layer model of
QPOs. As the accretion rate increases, material from the disk puffs up and
obscures the NS because the disk scale height increases (e.g., $H/R\sim
1$). Sco X-1 then enters the normal branch. X-ray photons are softened by the
added matter along the line-of-sight. This softening is observable as the
normal branch of the Z.

We have shown that all predicted frequencies of the transition layer model
for QPOs correspond with Sco X-1 observations. Furthermore, we have
identified the $\sim 6$ Hz frequencies in the normal branch as acoustic oscillations
of a spherical shell around the NS. This spherical shell is formed when the radiation
pressure near the Eddington accretion rate destroys the disk.
The size of the shell is on the order of one NS radii. We argue that there
is switching from disk accretion to spherical accretion in the vicinity of
the NS surface when the luminosity approaches Eddington.
Our confidence in this phenomenon for Sco X-1 is based on two tests
related to the QPOs classification picture and the hardness-intensity
diagram.
We view the striking correspondence between the data (Figures
\ref{fig1} \& \ref{fig2}) and the predictions
of the transition layer model as confirmation that we have correctly
identified the QPO phenomena observed in Sco X-1.
In addition, the best-fit parameters extracted from the photon spectra
(e.g., the plasma temperature and the optical depth of the transition
layer) enabled us to determine an upper limit of Sco X-1's transition
layer $B-$field while in the horizontal branch.


We thank Jean Swank and the staff of NASA's RXTE
Flight Operations Team and Guest Observer Facility at GSFC
 for X-ray data reduction support.
We also wish to thank Sergey Kuznetsov and Michael van der Klis for the
Sco X-1 data files.

\clearpage

\clearpage

\end{document}